\documentclass[letterpaper]{article}
\usepackage[preprint]{aaai2027}
\usepackage[hyphens]{url}
\usepackage{graphicx}
\usepackage{natbib}
\usepackage{caption}
\usepackage{algorithm}
\usepackage{algorithmic}
\usepackage{amsmath}
\usepackage{amssymb}
\usepackage{multirow}
\usepackage{booktabs}
\usepackage{arydshln}
\expandafter\let\csname endtabular*\endcsname\endtabular

\usepackage{newfloat}
\usepackage{listings}
\DeclareCaptionStyle{ruled}{labelfont=normalfont,labelsep=colon,strut=off}
\floatstyle{ruled}
\newfloat{listing}{tb}{lst}{}
\floatname{listing}{Listing}

\usepackage{booktabs}

\title{MSVAE: Rethinking Musical Semantics in Continuous Latent Spaces}
\author{
    Guobin~Ma\textsuperscript{\rm 1,\rm 2},
    Yuepeng~Jiang\textsuperscript{\rm 1},
    Dake~Guo\textsuperscript{\rm 1},
    Huakang~Chen\textsuperscript{\rm 1},
    Yuan~Zhou\textsuperscript{\rm 3},
    Lingxin~Xiong\textsuperscript{\rm 2},
    Wenkai~Cheng\textsuperscript{\rm 1},
    Jingbin~Hu\textsuperscript{\rm 1},
    Hanke~Xie\textsuperscript{\rm 1},
    Chunbo~Hao\textsuperscript{\rm 1},
    Wenhao~Li\textsuperscript{\rm 1},
    Ziqian~Ning\textsuperscript{\rm 1},
    Jixun~Yao\textsuperscript{\rm 1},
    Jian~Zhao\textsuperscript{\rm 2},
    Zhonglin~Jiang\textsuperscript{\rm 2},
    Yong~Chen\textsuperscript{\rm 2},
    Lei~Xie\textsuperscript{\rm 1},
    Pengcheng~Zhu\textsuperscript{\rm 4,\rm 5}
}
\affiliations{
    \textsuperscript{\rm 1}Audio, Speech and Language Processing Group (ASLP@NPU),\\
    School of Software, Northwestern Polytechnical University, China\\
    \textsuperscript{\rm 2}Geely Automobile Research Institute (Ningbo) Company Ltd, China\\
    \textsuperscript{\rm 3}Independent Researcher, China\\
    \textsuperscript{\rm 4}Faculty of Electronic and Information Engineering, Xi'an Jiaotong University, China\\
    \textsuperscript{\rm 5}WeNet Open Source Community, China\\
    guobin.ma@mail.nwpu.edu.cn, lxie@nwpu.edu.cn, zpcoftts@gmail.com
}

\newcommand{\MSVAEAppendixLayout}{\captionsetup{font=small,skip=4pt}\raggedbottom}

\begin{document}

\maketitle

\begin{abstract}
Continuous latent representations underlie recent diffusion-based lyrics-to-song and text-to-music systems, yet aligning lyric conditions with these latents remains difficult: existing systems resort to generator-side machinery such as sentence-level timestamps, auxiliary alignment objectives, or semi-autoregressive decoding, while the latents they predict are still produced by autoencoders trained purely for reconstruction. We address this difficulty from the representation side. We propose MSVAE (Music Semantic Variational Autoencoder)\footnote{Demo, code, and weights: \url{https://aslp-lab.github.io/MSVAE/}}, which structures its continuous latent around two complementary semantic streams---work-level compositional content and lyric-level linguistic-vocal content. Partitioned dual-stream alignment makes both streams explicit by aligning disjoint latent subspaces with frozen lead-sheet and ASR references, while a semantic-first gated residual posterior derives its base from a frozen music-SSL encoder and admits acoustic information only through gated, bounded corrections. MSVAE achieves competitive stereo reconstruction and stronger MARBLE probing than continuous-autoencoder baselines, and improves downstream lyrics-to-song and text-to-music generation. Notably, with only an untimed phoneme prefix and no lyric timing, MSVAE yields the lowest phoneme error rate, whereas all baseline representations require sentence-level alignment to produce intelligible vocals.

\end{abstract}

\section{Introduction}

Continuous latent representations have become a mainstream modeling space for recent diffusion-based generative models, including music generation models. 
Musical semantics encompass high-level melodic and harmonic organization as well as more explicit content relations such as lyric--vocal correspondence. Yet current continuous-latent systems struggle even with the latter: despite being directly observable and measurable, correspondence between lyric conditions and continuous latent sequences remains difficult to learn.
Existing systems address this problem through auxiliary representation-alignment objectives, sentence-level temporal supervision, post-training rewards for alignment, and semi-autoregressive decoding~\cite{DiffRhythm,ACESTEP,ACESTEP15,DiffRhythm2}. 
Despite their differences, all of these mechanisms are improvements on model design, while the target latents are mostly produced by autoencoders trained primarily for waveform reconstruction.

Purely reconstruction-oriented training of music autoencoders preserves the rich acoustic detail required for waveform synthesis, but nothing in it ensures that content factors such as melody, harmony, and lyric-related linguistic information are organized in an explicit and predictable way in the latent space.
As a result, the downstream generator must learn not only the latent distribution but also how these high-level content factors are organized within an acoustically oriented space, transferring substantial content-modeling pressure to the generator. 
Consistent with this, recent work in speech and general audio synthesis \cite{SemanticVAE,DashengTokenizer,WavCube,LoSATok,MingUniAudio,HoliTok} uses semantic supervision or self-supervised encoder features to learn continuous audio latents. These studies show that semantic guidance can encourage more structured latent spaces that are better suited to downstream generation.
This motivates addressing the difficulty from the representation side — designing a music semantic VAE whose latent supports high-fidelity reconstruction while explicitly structuring musical content. 
This raises a fundamental question: \emph{what constitutes musical semantics?}

Song-level semantic content comprises two primary, complementary streams: work-level compositional content, including melody and harmony, and lyric-level linguistic-vocal content~\cite{DBLP:conf/ismir/Meseguer-Brocal18}. 
The two semantic streams do not correspond to separate audio sources — vocal melody itself belongs to compositional content: a sung phrase carries melodic contour while simultaneously conveying phonetic content.
ACE-Step reflects this distinction in practice: its REPA~\cite{REPA} objective separately aligns the same DiT representation with MERT~\cite{MERT} and mHuBERT~\cite{mHuBERT}, drawing music- and speech-oriented semantic guidance in parallel. Yet the mHuBERT component is reduced to 0.01 during fine-tuning to avoid suppressing instrumental sounds~\cite{ACESTEP}, indicating that heterogeneous semantic objectives can interfere when imposed on an undivided representation. 
The first challenge is therefore to make both streams explicit within a compact latent without letting them interfere with each other.

An SSL encoder does not explicitly distinguish the two streams defined above, but its output is already organized around content and therefore provides a more structured substrate for making the two streams explicit. At the same time, SSL features do not preserve all the acoustic information required for high-fidelity reconstruction. This motivates combining SSL-derived semantic structure with reconstruction-relevant acoustic information in a shared continuous latent. Without appropriate constraints, however, the reconstruction objective may dominate this fusion and make the latent depend primarily on the acoustic pathway, thereby weakening its semantic organization. 
The second challenge is therefore to construct a continuous music latent that preserves a clear and predictable semantic organization while retaining the acoustic detail required for high-fidelity reconstruction.

To address these challenges, we propose \textbf{MSVAE} (Music Semantic Variational Autoencoder), built on partitioned dual-stream alignment and a semantic-first gated residual posterior. The partitioned objective aligns disjoint latent subspaces with representations from frozen SheetSage~\cite{SheetSage} and FireRedASR~\cite{FireRedASR} encoders, making compositional and linguistic-vocal content explicit while limiting interference between their supervision signals. The gated residual posterior derives its semantic base from a frozen MuQ~\cite{MuQ} encoder and admits reconstruction-relevant acoustic information only through bounded residual corrections. MSVAE embeds explicit content organization within the VAE latent rather than leaving it entirely to the downstream generator.

The main contributions are summarized as follows:
\begin{itemize}
    \item We propose MSVAE, a Music Semantic VAE that addresses lyric--audio correspondence at the representation level by organizing compositional and linguistic-vocal content in separate latent subspaces.
    
    \item We introduce a semantic-first gated residual posterior, which derives its semantic substrate from frozen MuQ features and integrates reconstruction-relevant acoustic details via bounded residual corrections.
    
    \item MSVAE demonstrates competitive reconstruction, superior MARBLE probing, and improved downstream lyrics-to-song and text-to-music generation. Notably, it enables intelligible vocal synthesis from untimed phoneme prefixes without relying on sentence-level lyric timing.
\end{itemize}

\section{Related Work}

\noindent\textbf{Lyric alignment in song generation.}
Continuous-latent song generators have introduced a range of generator-side mechanisms for lyric alignment. DiffRhythm~\cite{DiffRhythm} uses sentence-level timestamps during both training and inference. ACE-Step~\cite{ACESTEP} adds REPA~\cite{REPA} objectives that align intermediate DiT features with MERT and mHuBERT representations, while ACE-Step 1.5~\cite{ACESTEP15} further applies reinforcement learning with an intrinsic reward derived from the consistency of its lyric–frame cross-attention maps. DiffRhythm 2~\cite{DiffRhythm2} adopts semi-autoregressive block flow matching, preserving parallel generation within each block while introducing autoregressive progression across blocks and thus giving up full-sequence parallelism. These mechanisms operate at the generator level and leave the autoencoder latent unchanged; MSVAE instead addresses lyric–audio correspondence at the representation level.

\noindent\textbf{Music-oriented continuous representations.}
Continuous representations for music generation have primarily served as high-fidelity reconstruction interfaces. Stable Audio Open~\cite{StableAudioOpen} and ACE-Step 1.5~\cite{ACESTEP15} employ convolutional VAEs optimized mainly for acoustic reconstruction, without imposing explicit semantic structure on the latent. SAME~\cite{SAME} introduces semantic regularization into a Transformer-based music autoencoder, but its training data contain no vocal songs, and its objectives do not explicitly distinguish melodic--harmonic from lyric-related linguistic-vocal content. How to structure a reconstructive continuous latent around these music-specific semantic streams therefore remains unresolved.

\begin{figure*}[t]
  \centering
  \includegraphics[width=\textwidth]{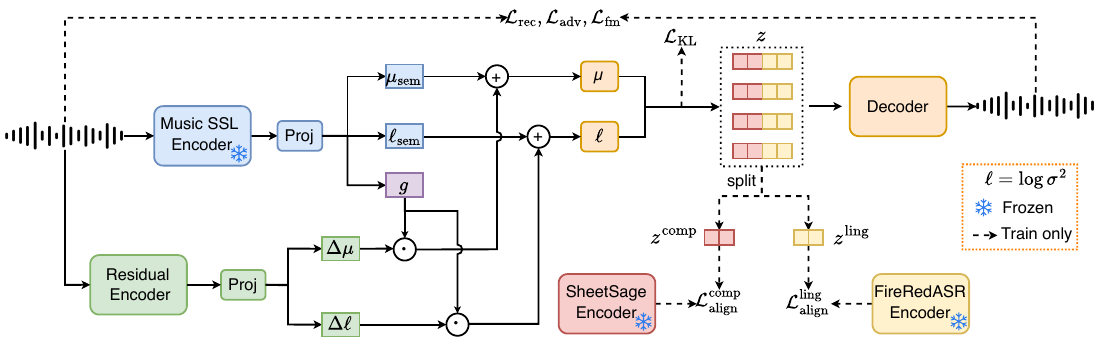}
  \caption{Overall architecture of our proposed MSVAE.}
  \label{fig:overview}
  \vspace{-13pt}
\end{figure*}

\section{Methodology}

\subsection{Overview}
\label{sec:method_overview}

Given a stereo music waveform $x\in\mathbb{R}^{2\times L}$, MSVAE encodes it into a 64-dimensional stochastic continuous representation $z\in\mathbb{R}^{T\times64}$ at 25~Hz. As illustrated in Fig.~\ref{fig:overview}, MSVAE combines partitioned dual-stream semantic alignment with a semantic-first gated residual posterior. The posterior derives its semantic base from features extracted by a frozen MuQ encoder and admits information from a trainable acoustic encoder only through bounded residual corrections whose contributions are gated by the projected MuQ features. A sample from the resulting posterior is decoded to reconstruct the input waveform. During training, $z$ is split into two 32-dimensional partitions, which are projected and aligned with representations from frozen SheetSage and FireRedASR encoders, corresponding to compositional melodic--harmonic and lyric-related linguistic-vocal content, respectively. The model is jointly trained with reconstruction and adversarial losses, KL regularization, and two stream-specific alignment objectives.

\subsection{Partitioned Dual-Stream Semantic Alignment}
\label{sec:partitioned_alignment}

A complete song combines a singing voice with polyphonic accompaniment. Across these components, melody and harmony characterize composition-level musical content, while singing additionally carries phonetic and lexical information. We therefore operationalize song-level semantics as two complementary streams: compositional melodic--harmonic content and lyric-related linguistic-vocal content. This factorization concerns semantic content rather than isolated audio sources: a sung melody contributes to the compositional stream while simultaneously conveying linguistic content.

To construct the compositional reference, we use hidden states from SheetSage's separately trained melody-transcription and chord-recognition Transformers \cite{SheetSage}. SheetSage is an audio-to-lead-sheet transcription system trained on approximately 50 hours of aligned melody and harmony annotations. Let
\begin{equation}
H^{\mathrm{melody}}
\in\mathbb{R}^{N\times d_{\mathrm{melody}}},
\qquad
H^{\mathrm{harmony}}
\in\mathbb{R}^{N\times d_{\mathrm{harmony}}}
\end{equation}
denote its melody and harmony hidden sequences, respectively, where $d_{\mathrm{melody}}+d_{\mathrm{harmony}}=1024$ in our implementation. 
Because these representations are indexed by beat-synchronous positions rather than a fixed frame rate, we expand them to the 25-Hz VAE grid by repeating each hidden vector in proportion to its duration, compensating for rounding errors to preserve sequence length. The expanded sequences are concatenated along the feature dimension to form the compositional reference
\begin{equation}
H^{\mathrm{comp}}
=
[\widetilde{H}^{\mathrm{harmony}}
\Vert
\widetilde{H}^{\mathrm{melody}}]
\in\mathbb{R}^{T\times1024}.
\label{eq:composition_reference}
\end{equation}

For the linguistic-vocal stream, we use the final encoder hidden states of a frozen FireRedASR-AED-L model \cite{FireRedASR}. FireRedASR-AED-L uses a 16-layer Conformer encoder with a model width of 1280, followed by a Transformer decoder, and is trained for speech and singing transcription. Two stride-2 convolutional layers reduce its 10-ms input feature sequence to a nominal 40-ms frame interval. The resulting encoder representation
\begin{equation}
H^{\mathrm{ling}}
\in\mathbb{R}^{T\times1280}
\label{eq:linguistic_reference}
\end{equation}
therefore operates at approximately 25~Hz, allowing the corresponding segment to be selected without temporal interpolation. We use these transcription-oriented hidden states as an operational reference for the lyric-related linguistic content of singing.

The reference sequences above use single-example notation. During training, they are stacked along the batch dimension as
$H^{\mathrm{comp}}\in\mathbb{R}^{B\times T\times1024}$ and
$H^{\mathrm{ling}}\in\mathbb{R}^{B\times T\times1280}$.
Given the sampled latent $z\in\mathbb{R}^{B\times T\times64}$, we partition it along the channel dimension as
\begin{equation}
z
=
[z^{\mathrm{comp}}\Vert z^{\mathrm{ling}}],
\qquad
z^{\mathrm{comp}},z^{\mathrm{ling}}
\in\mathbb{R}^{B\times T\times32}.
\label{eq:latent_partition}
\end{equation}
Two independent frame-wise projection heads map the latent partitions to the dimensions of their corresponding reference representations:
\begin{equation}
\begin{aligned}
A^{\mathrm{comp}}
&=
P_{\mathrm{comp}}(z^{\mathrm{comp}})
\in\mathbb{R}^{B\times T\times1024},\\
A^{\mathrm{ling}}
&=
P_{\mathrm{ling}}(z^{\mathrm{ling}})
\in\mathbb{R}^{B\times T\times1280}.
\end{aligned}
\label{eq:stream_projection}
\end{equation}
Only the latent partitions are projected; the precomputed reference representations remain unprojected. The reference encoders are used only for offline feature extraction, while the projection heads are required only during VAE training.

We then align each projected partition with its temporally matched reference using a feature-axis objective. For $A,H\in\mathbb{R}^{B\times T\times d}$, the loss is
\begin{equation}
\ell_{\mathrm{axis}}(A,H)
=
-\frac{1}{BT}\sum_{b,t}
\log\operatorname{sigmoid}
\left(\cos(A_{b,t},H_{b,t})\right),
\label{eq:axis_alignment}
\end{equation}
where $\operatorname{cos}(\cdot,\cdot)$ denotes cosine similarity along the feature dimension.
The two stream-specific alignment objectives are
\begin{equation}
\begin{aligned}
\mathcal{L}_{\mathrm{align}}^{\mathrm{comp}}
&=
\ell_{\mathrm{axis}}
\left(A^{\mathrm{comp}},H^{\mathrm{comp}}\right),\\
\mathcal{L}_{\mathrm{align}}^{\mathrm{ling}}
&=
\ell_{\mathrm{axis}}
\left(A^{\mathrm{ling}},H^{\mathrm{ling}}\right).
\end{aligned}
\label{eq:partitioned_alignment}
\end{equation}

A full-dimensional alternative aligns both references with all 64 latent channels. Our partitioned design instead assigns them to disjoint 32-dimensional channel sets, limiting interference between the two alignment objectives and encouraging complementary specialization. We compare the two designs in Table~\ref{tab:ablation}.

\subsection{Semantic-First Gated Residual Posterior}
\label{sec:gated_residual_posterior}
Rather than symmetrically fusing SSL and acoustic features, MSVAE uses MuQ-derived parameters as the base of its latent distribution and admits acoustic information only through gated, bounded corrections. Throughout, a diagonal Gaussian is parameterized by its mean $\mu$ and log-variance $\ell=\log\sigma^{2}$. The semantic base parameters are constructed from the sixth-layer representation of a frozen MuQ encoder:
\begin{equation}
\begin{aligned}
e_{\mathrm{sem}} &= P_{\mathrm{sem}}\!\left(E_{\mathrm{MuQ}}^{(6)}(x)\right),\\
\mu_{\mathrm{sem}} &= f_{\mu}^{\mathrm{sem}}(e_{\mathrm{sem}}),\\
\ell_{\mathrm{sem}} &= f_{\ell}^{\mathrm{sem}}(e_{\mathrm{sem}}).
\end{aligned}
\label{eq:semantic_base_parameters}
\end{equation}
The MuQ encoder is frozen, while $P_{\mathrm{sem}}$ and the two parameter heads are trained end-to-end with the rest of the model.

For $r\in\{\mu,\ell\}$, the acoustic residual encoder $E_{\mathrm{res}}$ predicts a bounded correction:
\begin{equation}
\begin{aligned}
s_r &= s_{\max}\operatorname{sigmoid}(\alpha_r),\\
\Delta r &= s_r\tanh\left(f_r^{\mathrm{res}}\!\left(E_{\mathrm{res}}(x)\right)\right),
\end{aligned}
\label{eq:bounded_residuals}
\end{equation}
where $\alpha_r$ is learned and $s_{\max}$ is a fixed upper bound; the acoustic branch thus predicts bounded corrections rather than an independent latent distribution. A gate computed from the projected MuQ feature determines how much of each correction enters the final distribution:
\begin{equation}
\begin{aligned}
g &= \operatorname{sigmoid}\left(f_g(e_{\mathrm{sem}})\right),\\
\mu &= \mu_{\mathrm{sem}}+g\odot\Delta\mu,\\
\ell &= \ell_{\mathrm{sem}}+g\odot\Delta\ell.
\end{aligned}
\label{eq:gated_residual_posterior}
\end{equation}
Since $0<g<1$ and the residual predictions are bounded by $\tanh$, each acoustic correction is bounded by $s_{\max}$, and conditioning $g$ on $e_{\mathrm{sem}}$ makes the admission of acoustic information depend on the semantic substrate itself. The MuQ-derived parameters therefore remain the explicit base of the final distribution, without an unconstrained acoustic pathway that can replace them.

Let $\sigma=\exp(\ell/2)$; the approximate posterior $q_{\phi}(z\mid x)=\mathcal{N}\left(\mu,\operatorname{diag}(\sigma^{2})\right)$ is sampled via reparameterization:
\begin{equation}
z = \mu+\sigma\odot\epsilon,\qquad \epsilon \sim \mathcal{N}(0,I).
\label{eq:posterior_sampling}
\end{equation}
The sampled $z$ is used for both waveform reconstruction and the two partitioned alignment objectives.

\subsection{Training Objectives}
\label{sec:training_objectives}

Let $\hat{x}=\operatorname{Dec}_{\theta}(z)$ denote the stereo waveform reconstructed from the sampled latent. The trainable components of MSVAE are optimized jointly, while the MuQ, SheetSage, and FireRedASR encoders remain frozen. The reconstruction objective combines a multi-resolution STFT loss \cite{ParallelWaveGAN} with a stereo sum-and-difference STFT loss \cite{StereoSTFT}:
\begin{equation}
\begin{split}
\mathcal{L}_{\mathrm{rec}}
&=
\lambda_{\mathrm{stft}}
\mathcal{L}_{\mathrm{MR\text{-}STFT}}(x,\hat{x})\\
&\quad+
\lambda_{\mathrm{stereo}}
\mathcal{L}_{\mathrm{SD\text{-}STFT}}(x,\hat{x}).
\end{split}
\label{eq:reconstruction_loss}
\end{equation}
The multi-resolution STFT term is evaluated independently on the left and right channels and then summed, whereas the stereo term is computed from their sum and difference signals.

To complement the spectral reconstruction objectives, we employ the multi-period discriminator (MPD) and multi-resolution discriminator (MRD) used in DAC \cite{DAC}, together with an MSS-CQT discriminator \cite{MSSBCQT}. We aggregate the generator-side adversarial and feature-matching losses produced by these discriminator families and denote them by $\mathcal{L}_{\mathrm{adv}}$ and $\mathcal{L}_{\mathrm{fm}}$, respectively. Relative discriminator-specific weights are absorbed into these aggregate terms, while $\lambda_{\mathrm{adv}}$ and $\lambda_{\mathrm{fm}}$ control their overall contributions. The discriminators are optimized using their corresponding standard adversarial objectives, whose aggregate is denoted by $\mathcal{L}_{D}$.

Consistent with the latent partition used for semantic alignment, we regularize the corresponding marginals of the final gated residual posterior separately. Let $\mathcal{S}=\{\mathrm{comp},\mathrm{ling}\}$ index the two semantic streams, and let $(\mu^{s},\ell^{s})$ denote the corresponding 32-dimensional partitions of the final posterior parameters. For each $s\in\mathcal{S}$, $q_{\phi}(z_{b,t}^{s}\mid x)$ denotes the diagonal-Gaussian marginal parameterized by $\mu_{b,t}^{s}$ and $\ell_{b,t}^{s}$, and $p_{0}=\mathcal{N}(0,I_{32})$ denotes the standard normal prior. The corresponding KL objective is
\begin{equation}
\mathcal{L}_{\mathrm{KL}}^{s}
=
\frac{1}{BT}
\sum_{b,t}
D_{\mathrm{KL}}
\left(
q_{\phi}(z_{b,t}^{s}\mid x)
\Vert p_{0}
\right).
\label{eq:partitioned_kl}
\end{equation}
Computing the two KL terms separately allows their regularization strengths to be controlled independently.

Combining the reconstruction, adversarial, feature-matching, variational, and semantic alignment terms, the complete generator-side objective is
\begin{equation}
\begin{aligned}
\mathcal{L}_{G} ={}& \mathcal{L}_{\mathrm{rec}}+\lambda_{\mathrm{adv}}\mathcal{L}_{\mathrm{adv}}+\lambda_{\mathrm{fm}}\mathcal{L}_{\mathrm{fm}}\\
&+ \sum_{s\in\mathcal{S}}\left(\lambda_{\mathrm{KL}}^{s}\mathcal{L}_{\mathrm{KL}}^{s}+\lambda_{\mathrm{align}}^{s}\mathcal{L}_{\mathrm{align}}^{s}\right).
\end{aligned}
\label{eq:generator_objective}
\end{equation}
The trainable MSVAE components are updated using $\mathcal{L}_{G}$, while the discriminator parameters are optimized separately using $\mathcal{L}_{D}$. This joint formulation imposes acoustic fidelity, prior regularization, and dual-stream semantic alignment on the same sampled latent.

\section{Experiments}

\subsection{Experimental Setup}

\noindent\textbf{Training data and schedule.}
MSVAE is trained for 500k steps on approximately 600k vocal and instrumental tracks totaling about 40k hours, resampled to 24~kHz stereo, using random 4-second crops and a batch size of 32. We then freeze all other components and refine only the decoder for another 500k steps on approximately 300k tracks totaling about 20k hours, using 48-kHz reconstruction targets.

\noindent\textbf{Model configuration.}
MSVAE produces a 64-dimensional, 25-Hz latent using a DAC-style convolutional residual encoder~\cite{DAC} with $960\times$ downsampling and a BigVGAN-style decoder~\cite{BigVGAN}. MuQ processes the two stereo channels independently, and the MuQ, SheetSage, and FireRedASR encoders remain frozen. Complete implementation details are provided in the Appendix.

\noindent\textbf{Baselines and comparison protocol.}
We compare official checkpoints of Stable Audio Open~\cite{StableAudioOpen} and ACE-Step 1.5~\cite{ACESTEP15}, both convolutional VAEs, and the Transformer-based, semantically regularized SAME-S/L~\cite{SAME}. Within each downstream task, all representations share the training data, generator architecture and scale, optimization schedule, and number of training steps; only the input/output projections and sequence lengths are adapted to their latent dimensions and frame rates.

\begin{table}[t]
\centering
\resizebox{\linewidth}{!}{%
\fontsize{8.5pt}{10.2pt}\selectfont
\setlength{\tabcolsep}{3pt}
\renewcommand{\arraystretch}{1.06}
\begin{tabular}{@{}lcccccc@{}}
\toprule
Model & Latent & Params & PESQ$\uparrow$ & STOI$\uparrow$ & STFT$_{\log}\downarrow$ & Mel$_{\log}\downarrow$ \\
\midrule
\multicolumn{7}{@{}l}{\textit{64-dimensional convolutional VAEs}} \\
Stable Audio Open
& $64{\times}21.5$ & 156M
& 1.784 & 0.699 & 0.139 & 0.101 \\
ACE-Step 1.5
& $64{\times}25$ & 169M
& 1.912 & \textbf{0.756} & 0.137 & 0.092 \\
\textbf{MSVAE}
& $64{\times}25$ & 193M
& \textbf{2.253} & 0.740 & \textbf{0.136} & \textbf{0.090} \\
\midrule
\multicolumn{7}{@{}l}{\textit{256-dimensional Transformer autoencoders}} \\
SAME-S
& $256{\times}10.8$ & 108M
& 2.414 & 0.782 & 0.131 & 0.085 \\
SAME-L
& $256{\times}10.8$ & 852M
& \textbf{2.886} & \textbf{0.840} & \textbf{0.121} & \textbf{0.068} \\
\bottomrule
\end{tabular}%
}
\caption{Stereo music reconstruction at a common 24-kHz evaluation rate. Models are grouped by trainable backbone family and latent dimensionality. The latent column reports dimensionality $\times$ frame rate in Hz, and bold denotes the best result within each block. Parameter counts include trainable components only.}
\vspace{-10pt}
\label{tab:reconstruction}
\end{table}

\subsection{Music Reconstruction}

\subsubsection{Setup}
We evaluate music reconstruction on the official 50-track test split of MUSDB18-HQ~\cite{musdb18hq}. We report four complementary metrics: PESQ~\cite{PESQ} and STOI~\cite{STOI} emphasize perceptual quality and intelligibility~\cite{QwenAudioVAE}, whereas $\mathrm{STFT}_{\log}$ and $\mathrm{Mel}_{\log}$ measure spectral fidelity following the multi-resolution definitions of SAME~\cite{SAME}. PESQ is computed in wideband mode at 16~kHz and STOI at 10~kHz; all metrics are computed per channel and averaged.

\subsubsection{Results}
Within the 64-dimensional convolutional VAE block in Table~\ref{tab:reconstruction}, MSVAE achieves the best PESQ and spectral distances, while its STOI is second only to ACE-Step 1.5. Explicit semantic organization is therefore compatible with competitive acoustic fidelity at a comparable latent and trainable-parameter budget.
The SAME variants attain higher reconstruction scores but use a $1.7\times$ larger latent-rate budget. Both also undergo two encoder-frozen decoder-finetuning stages designed to improve reconstruction quality; SAME-L is substantially larger, while SAME-S is distilled from SAME-L.

\subsection{Music Understanding}

\subsubsection{Setup}
We evaluate music understanding on selected tasks from the MARBLE benchmark~\cite{marble}. All representation models are kept frozen, and only task-specific probing heads are trained under identical data splits and optimization settings. The selected tasks cover complementary aspects of musical content: EMO evaluates music emotion regression through the coefficients of determination $R^2_{\mathrm{V}}$ and $R^2_{\mathrm{A}}$ for valence and arousal; Chords evaluates large-vocabulary chord recognition with accuracy; GS evaluates musical key detection on GiantSteps with weighted accuracy; and GTZAN Beat evaluates beat tracking with F1 score.

\subsubsection{Results}
Table~\ref{tab:marble} shows that MSVAE outperforms all continuous-autoencoder baselines across the five reported metrics. The largest gain appears in valence regression, where $R_V^2$ increases from $0.267$ to $0.469$; clear gains are also observed in chord recognition and beat tracking. Despite using only a 64-dimensional latent at 25~Hz and additionally supporting waveform decoding, MSVAE surpasses the 1024-dimensional SheetSage reference on four of the five metrics and MuQ layer~6 on key estimation. These results demonstrate that MSVAE preserves substantial musical semantic information within a compact, reconstructive latent representation.

\begin{table}[t]
\centering
\small
\setlength{\tabcolsep}{1.5pt}
\renewcommand{\arraystretch}{1.08}
\begin{tabular}{@{}lcccccc@{}}
\toprule
\multirow[c]{2}{*}{Model} & \multirow[c]{2}{*}{Dim} & \multicolumn{2}{c}{EMO} & \multicolumn{1}{c}{Chords} & \multicolumn{1}{c}{GS} & \multicolumn{1}{c}{GTZAN} \\
\cmidrule(lr){3-4}\cmidrule(lr){5-5}\cmidrule(lr){6-6}\cmidrule(lr){7-7}
& & $R^2_{\mathrm{V}}\,\uparrow$ & $R^2_{\mathrm{A}}\,\uparrow$ & Acc.\,$\uparrow$ & Acc.\,$\uparrow$ & Beat F1\,$\uparrow$ \\
\midrule
\multicolumn{7}{@{}l}{\textit{Frozen references (no decoder)}} \\
SheetSage   & 1024 & 0.344 & 0.614 & 0.698 & 0.623 & 0.356 \\
MuQ layer 6 & 1024 & 0.579 & 0.739 & 0.714 & 0.561 & 0.744 \\
\midrule
\multicolumn{7}{@{}l}{\textit{Continuous autoencoders}} \\
Stable Audio Open & 64  & 0.161 & 0.572 & 0.377 & 0.494 & 0.389 \\
ACE-Step 1.5 & 64  & \underline{0.267} & \underline{0.578} & 0.383 & 0.567 & 0.389 \\
SAME-S           & 256 & 0.165 & 0.526 & 0.531 & \underline{0.609} & \underline{0.506} \\
SAME-L           & 256 & 0.199 & 0.532 & \underline{0.548} & 0.599 & 0.503 \\
\midrule
\textbf{MSVAE}   & 64  & \textbf{0.469} & \textbf{0.672} & \textbf{0.634} & \textbf{0.634} & \textbf{0.576} \\
\bottomrule
\end{tabular}
\caption{Linear-probing results on selected MARBLE tasks. Frozen encoder features are shown for context and excluded from the comparison among continuous autoencoders.}
\vspace{-10pt}
\label{tab:marble}
\end{table}

\begin{table*}[t]
\centering
\small
\setlength{\tabcolsep}{3.5pt}
\renewcommand{\arraystretch}{1.06}
\begin{tabular}{@{}llcccccccccccc@{}}
\toprule
\multirow[c]{2}{*}{Lyric cond.} &
\multirow[c]{2}{*}{Model} &
\multicolumn{9}{c}{Lyrics-to-Song} &
\multicolumn{3}{c}{Text-to-Music} \\
\cmidrule(lr){3-11}\cmidrule(lr){12-14}
& & PQ$\uparrow$ & CE$\uparrow$ & Coh.$\uparrow$ & Mus.$\uparrow$
& Mem.$\uparrow$ & Cla.$\uparrow$ & Nat.$\uparrow$
& MuLan$\uparrow$ & PER$\downarrow$
& FAD$\downarrow$ & CLAP$\uparrow$ & CCS$\uparrow$ \\
\midrule
\multirow{5}{*}{\shortstack[l]{Sentence-level\\timestamps}}
& Stable Audio Open
& 7.33 & 6.80 & 2.45 & 2.18 & 2.22 & 2.20 & 2.18
& 0.467 & 0.366 & 0.677 & 0.135 & 0.650 \\

& ACE-Step 1.5
& 7.90 & \underline{7.42} & 3.05 & 2.83 & 2.83 & 2.82 & 2.70
& 0.618 & 0.141 & 0.687 & 0.171 & \underline{0.698} \\

& SAME-S
& 7.57 & 7.13 & 2.72 & 2.51 & 2.57 & 2.53 & 2.42
& 0.576 & 0.344 & 0.675 & \underline{0.179} & 0.635 \\

& SAME-L
& 7.68 & 7.16 & 2.87 & 2.65 & 2.70 & 2.63 & 2.60
& 0.570 & 0.290 & \underline{0.660} & \textbf{0.180} & 0.647 \\

& \textbf{MSVAE}
& \textbf{8.03} & 7.25
& \underline{3.37} & \underline{3.15} & \underline{3.29}
& \underline{3.23} & \underline{3.11} & \underline{0.646}
& \textbf{0.069}
& --- & --- & --- \\
\midrule
Untimed prefix
& \textbf{MSVAE}
& \underline{8.02} & \textbf{7.44}
& \textbf{3.60} & \textbf{3.39} & \textbf{3.46}
& \textbf{3.38} & \textbf{3.26} & \textbf{0.678}
& \underline{0.136}
& \textbf{0.547} & 0.174 & \textbf{0.735} \\
\bottomrule
\end{tabular}
\caption{Downstream generation results grouped by lyric-conditioning protocol. Baselines require sentence-level timestamps; without them, PER exceeds $0.80$ (Appendix). MSVAE is evaluated with both sentence-level timestamps and an untimed phoneme prefix. Bold and underlined values denote the best and second-best results. Text-to-music is independent of lyric conditioning and is reported once for MSVAE.}
\label{tab:downstream_generation}
\end{table*}

\subsection{Downstream Music Generation}

\subsubsection{Setup}
We evaluate each representation as the modeling space of a downstream generative model. Within each task, all representations use the same generator backbone and model scale, training data, optimization schedule, and number of training steps.

For lyrics-to-song, we train a 1B non-autoregressive latent diffusion transformer following DiffRhythm+~\cite{DiffRhythmPlus} on 200k songs at a training length of 45 seconds, and evaluate on 200 AI-generated lyrics. The generators built on the baseline representations follow the sentence-level alignment paradigm of DiffRhythm~\cite{DiffRhythm}. For MSVAE, we train one generator with the same timestamped conditioning and another with an untimed phoneme-token prefix, the latter serving as our primary setting. Audiobox Aesthetics~\cite{AudioboxAesthetics} provides production quality (PQ) and content enjoyment (CE), while SongEval~\cite{SongEval} provides five song-level aesthetic dimensions. MuLan measures prompt--audio agreement in the MuQ-MuLan~\cite{MuQ} embedding space, and phoneme error rate (PER) measures lyric intelligibility, computed on vocals separated by YingMusic-Singer~\cite{YingMusicSinger} and transcribed with Qwen3-ASR~\cite{Qwen3ASR}.

For text-to-music, we follow the ICME2026 ATTM Grand Challenge protocol~\cite{attm}: a 400M conditional DiT based on DiffRhythm+ is trained on 30-second instrumental segments from MTG-Jamendo (3.7k hours)~\cite{MTGJamendo} and evaluated on the official test prompts. We report Fréchet audio distance (FAD)~\cite{FAD}, CLAP score for text--audio alignment, and concept coverage score (CCS), which uses a large audio-language model to verify the presence of specified musical concepts.

\subsubsection{Lyrics-to-Song}
As shown in Table~\ref{tab:downstream_generation}, the untimed MSVAE setting achieves the best CE, all five SongEval scores, and MuLan, with every SongEval dimension improving by more than $0.5$ points over the strongest baseline representation. Despite receiving no explicit lyric timing, it attains a PER of $0.136$, lower than every timestamp-conditioned baseline. Providing MSVAE with sentence-level timestamps further reduces PER to $0.069$, but lowers CE, all five SongEval dimensions, and MuLan, revealing a trade-off between stronger lyric alignment and song-level quality; we therefore retain the untimed configuration as our primary setting. In contrast, when sentence-level alignment is replaced with an untimed prefix, vocals generated from every baseline representation no longer follow the supplied lyrics under the shared generator recipe (PER $>0.80$; Appendix).

This timestamp-free capability extends beyond the 45-second setting. When trained from scratch at 90 and 150 seconds with the same untimed conditioning and no auxiliary generator-side alignment objective, MSVAE continues to produce intelligible vocals, with musicality remaining close to the 45-second setting although PER increases with duration (Appendix). These results indicate that the linguistic-vocal structure of the MSVAE latent substantially reduces reliance on externally specified lyric timing.

\subsubsection{Text-to-Music}
MSVAE also achieves the strongest overall performance on text-to-music generation, reducing FAD from the best baseline result of $0.660$ to $0.547$ and improving CCS from $0.698$ to $0.735$. These improvements indicate better distributional quality and more reliable realization of the musical concepts specified by the prompt. Its CLAP score remains competitive, although slightly below those of the two SAME variants. This is expected, as SAME-L is trained with a cross-modal contrastive objective that explicitly aligns its latent with caption embeddings, and SAME-S inherits this property through distillation from SAME-L. MSVAE attains its FAD and CCS improvements without an audio–caption alignment objective during VAE training.

\begin{table*}[t]
\centering
\small
\setlength{\tabcolsep}{4pt}
\renewcommand{\arraystretch}{1.06}
\begin{tabular}{@{}llccccccc@{}}
\toprule
\multirow[c]{2}{*}{Posterior design} &
\multirow[c]{2}{*}{Semantic alignment} &
\multicolumn{2}{c}{Reconstruction} &
\multicolumn{2}{c}{MARBLE probing} &
\multicolumn{3}{c}{Lyrics-to-Song} \\
\cmidrule(lr){3-4}\cmidrule(lr){5-6}\cmidrule(lr){7-9}
& & PESQ$\uparrow$ & Mel$_{\log}\downarrow$ & Chords Acc.$\uparrow$ & EMO $R^2_{\mathrm{V}}\uparrow$ & PQ$\uparrow$ & Mus.$\uparrow$ & PER$\downarrow$ \\
\midrule
\textbf{Gated residual (MSVAE)} & Lead+ASR (part.) & 2.253 & 0.090 & 0.634 & 0.469 & 8.02 & 3.39 & 0.136 \\
\midrule
\multicolumn{9}{@{}l}{\textit{Semantic alignment}} \\
Gated residual & None & 1.468 & 0.128 & 0.500 & 0.459 & 7.71 & 3.27 & 0.329 \\
Gated residual & Lead+ASR (full-dim.) & 1.451 & 0.128 & 0.619 & 0.517 & 7.81 & 3.45 & 0.341 \\
Gated residual & Lead only & 1.474 & 0.127 & 0.620 & 0.465 & 7.91 & 3.40 & 0.269 \\
Gated residual & ASR only & 1.472 & 0.127 & 0.471 & 0.512 & 7.87 & 3.39 & 0.236 \\
Gated residual & Chroma+ASR (part.) & 2.206 & 0.092 & 0.445 & 0.442 & 7.99 & 3.25 & 0.129 \\
\midrule
\multicolumn{9}{@{}l}{\textit{Posterior construction}} \\
Direct addition & Lead+ASR (part.) & 2.289 & 0.088 & 0.616 & 0.490 & 7.98 & 3.16 & 0.282 \\
Direct addition & None & 2.399 & 0.087 & 0.408 & 0.174 & 7.67 & 3.14 & 0.887 \\
MuQ only & Lead+ASR (part.) & 1.458 & 0.129 & 0.608 & 0.506 & 7.82 & 3.48 & 0.226 \\
MuQ only & None & 1.478 & 0.129 & 0.501 & 0.476 & 7.82 & 3.42 & 0.359 \\
\bottomrule
\end{tabular}
\caption{Ablation of semantic alignment and posterior construction. The upper block varies alignment under the gated residual posterior; the lower block compares direct-addition and MuQ-only posteriors, each with and without partitioned alignment. Lead and ASR denote SheetSage and FireRedASR; \emph{part.} assigns them to disjoint 32-dimensional subspaces, whereas \emph{full-dim.} aligns both with all 64 channels. Acoustic-only variants are reported in Table~\ref{tab:repa}.}
\vspace{-10pt}
\label{tab:ablation}
\end{table*}

\begin{table}[t]
\centering
\footnotesize
\setlength{\tabcolsep}{3pt}
\renewcommand{\arraystretch}{1.06}
\begin{tabular}{@{}lcccc@{}}
\toprule
\multirow[c]{2}{*}{Semantic guidance} &
\multicolumn{1}{c}{MARBLE} &
\multicolumn{3}{c}{Lyrics-to-Song} \\
\cmidrule(lr){2-2}\cmidrule(lr){3-5}
& EMO $R^2_{\mathrm{V}}\uparrow$ & PQ$\uparrow$ & Mus.$\uparrow$ & PER$\downarrow$ \\
\midrule
None & 0.152 & 7.50 & 3.09 & 1.394 \\
REPA (MERT+mHuBERT) & 0.152 & 7.93 & 3.08 & 0.694 \\
REPA (Lead+ASR) & 0.152 & 7.63 & 2.76 & 0.198 \\
Latent alignment (Lead+ASR) & \textbf{0.382} & \textbf{7.96} & \textbf{3.28} & \textbf{0.182} \\
\midrule
\emph{MSVAE (full; context)} & \emph{0.469} & \emph{8.02} & \emph{3.39} & \emph{0.136}  \\
\bottomrule
\end{tabular}
\caption{Representation-side versus generator-side semantic guidance under untimed lyric conditioning. The first three rows share an unaligned acoustic VAE; REPA modifies only generator training, leaving probe scores identical. Latent alignment retrains the VAE with partitioned Lead+ASR loss. Full MSVAE (italics) is excluded from bold ranking; see Appendix for details.}
\vspace{-5pt}
\label{tab:repa}
\end{table}

\subsection{Ablation Studies}
All lyrics-to-song ablations in this section use an untimed phoneme prefix without the sentence-level alignment paradigm.

\subsubsection{Semantic Alignment}
The upper block of Table~\ref{tab:ablation} varies the alignment strategy under a fixed gated residual posterior. Reconstruction exhibits the clearest pattern: PESQ remains near $1.47$ without alignment, under full-dimensional dual-stream alignment, or with either reference used alone, but rises to $2.253$ when the lead-sheet and ASR references are assigned to disjoint subspaces. Thus, neither dual-reference alignment without partitioning nor either semantic stream alone is sufficient; their partitioned combination is required for the gated residual posterior to attain high-fidelity reconstruction. It is also the only alignment variant that produces a substantial reconstruction gain while remaining strong in both probing and generation. Replacing the lead-sheet reference with chroma also unlocks reconstruction and retains a low PER, but degrades probing and song musicality, showing that effective subspace allocation also depends on the semantic content of the references.

\subsubsection{Posterior Design}
The lower block of Table~\ref{tab:ablation} compares alternative ways of constructing the posterior. Without alignment, direct addition achieves the strongest reconstruction (PESQ $2.399$), but its EMO $R^2_{\mathrm{V}}$ falls to $0.174$. By contrast, the unaligned gated residual posterior remains close to the MuQ-only variant ($0.459$ versus $0.476$), indicating that gating prevents the acoustic branch from overriding the semantics inherited from MuQ. Partitioned alignment substantially restores the probing performance of direct addition, yet its musicality and PER remain markedly worse than those of the gated residual posterior, making the benefit of gating most evident in downstream generation. The MuQ-only and acoustic-only posteriors expose complementary limitations: the former retains strong probing and musicality but remains near a PESQ of $1.47$, whereas the latter provides reconstruction comparable to the full model but leaves the latent weakly structured, with an EMO $R^2_{\mathrm{V}}$ of only $0.152$ in Table~\ref{tab:repa}. Complete reconstruction metrics for the acoustic-only variants are provided in the Appendix. By using MuQ as the posterior substrate and restricting the acoustic encoder to gated, bounded residual corrections, MSVAE combines these strengths, keeping reconstruction close to the strongest variants while attaining the best chord accuracy, PQ, and PER among the posterior-construction variants.

\subsubsection{Representation-Side versus Generator-Side Guidance}
Table~\ref{tab:repa} compares where semantic guidance should be introduced while fixing the posterior architecture to a single acoustic encoder and removing sentence-level lyric timing. Without guidance, the downstream generator fails to align lyrics, yielding a PER of $1.394$. Generator-side REPA improves intelligibility, but MERT+mHuBERT leaves musicality unchanged, while Lead+ASR reduces PER to $0.198$ at the cost of lowering musicality to $2.76$. Applying partitioned Lead+ASR alignment during VAE training instead achieves the best PQ, musicality, and PER among these controlled variants, while raising EMO $R^2_{\mathrm{V}}$ from $0.152$ to $0.382$. This comparison shows that structuring semantics in the latent provides a more effective modeling space than injecting semantic guidance only during generator training.

\subsubsection{Effect of the Linguistic Stream on Instrumental Music}
A natural concern is that the ASR-aligned stream may be detrimental when the target contains no linguistic-vocal content. Table~\ref{tab:ttm_ablation} shows the opposite: relative to lead-sheet-only alignment, the partitioned dual-stream configuration reduces FAD from $0.662$ to $0.547$ while attaining the best CLAP and CCS. Neither single-stream variant consistently improves over no alignment. A plausible explanation is that constraining both halves encourages compositional content to remain localized in its designated subspace. With lead-sheet-only alignment, the remaining dimensions receive no stream-specific target and may still absorb melodic--harmonic variation. Retaining the ASR-aligned stream may therefore benefit instrumental generation by completing the partition-wise organization rather than by providing active linguistic content.

\begin{table}[t]
\centering
\small
\setlength{\tabcolsep}{6pt}
\renewcommand{\arraystretch}{1.06}
\begin{tabular}{@{}lccc@{}}
\toprule
Semantic alignment & FAD\,$\downarrow$ & CLAP\,$\uparrow$ & CCS\,$\uparrow$ \\
\midrule
None                      & 0.679 & 0.165 & 0.692 \\
Lead only                 & 0.662 & 0.169 & 0.678 \\
ASR only                  & 0.688 & 0.158 & 0.686 \\
\textbf{Lead+ASR (part.)} & \textbf{0.547} & \textbf{0.174} & \textbf{0.735} \\
\bottomrule
\end{tabular}
\caption{Text-to-music ablation of semantic alignment on instrumental music. All variants use the gated residual posterior.}
\vspace{-5pt}
\label{tab:ttm_ablation}
\end{table}

\subsubsection{Subspace Specialization}
We probe the two latent partitions by zeroing one half of $z$ before decoding. Retaining only the compositional partition reconstructs the accompaniment and renders the vocal melody as instrument-like humming, but produces no intelligible lyrics. Conversely, retaining only the linguistic-vocal partition preserves intelligible singing while reducing the accompaniment to sparse rhythmic backing. These complementary interventions indicate that partitioned alignment induces distinct functional roles for the two latent subspaces.

\section{Conclusion}
We presented MSVAE, a Music Semantic VAE that structures its continuous latent around two complementary semantic streams---compositional melodic--harmonic content and lyric-related linguistic-vocal content---by combining partitioned dual-stream alignment with a semantic-first gated residual posterior. Although motivated by the lyric--audio correspondence problem, this representation-side design yields broader benefits: competitive reconstruction, stronger semantic probing, and improved downstream generation for both sung and instrumental music, including intelligible singing for up to 150 seconds without sentence-level lyric timing. These results highlight semantically structured continuous latents as a promising foundation for music generation.

\bibliography{aaai2027}

\clearpage
\MSVAEAppendixLayout
\section*{Appendix}

\appendix
\setcounter{secnumdepth}{2}
\setcounter{section}{0}
\renewcommand{\thesection}{\Alph{section}}
\renewcommand{\thesubsection}{\thesection.\arabic{subsection}}
\counterwithin{table}{section}
\counterwithin{figure}{section}

\renewcommand{\topfraction}{0.95}
\renewcommand{\bottomfraction}{0.95}
\renewcommand{\textfraction}{0.05}
\renewcommand{\floatpagefraction}{0.80}
\renewcommand{\dbltopfraction}{0.95}
\setcounter{dbltopnumber}{4}
\setlength{\dbltextfloatsep}{10pt plus 2pt minus 2pt}
\setlength{\dblfloatsep}{8pt plus 2pt minus 2pt}
\renewcommand{\dblfloatpagefraction}{0.80}

\section{Implementation Details}
\label{app:implementation}

\subsection{MSVAE Configuration}
\label{app:msvae_config}

Table~\ref{tab:msvae-config} summarizes the architecture and two-stage training configuration, and Table~\ref{tab:loss-weights} lists the MSVAE training-loss weights. Stage~II initializes from the Stage~I checkpoint and refines only the 48-kHz decoder, while the encoder remains frozen.

\par\addvspace{6pt}
\noindent\begin{minipage}{\columnwidth}
\setlength{\tabcolsep}{3pt}
\centering
\normalsize
\renewcommand{\arraystretch}{1.18}
\begin{tabular}{@{}lcc@{}}
\toprule
 & Stage~I & Stage~II \\
\midrule
I/O rate (kHz)     & 24\,/\,24 & 24\,/\,48 \\
Latent             & \multicolumn{2}{c}{$64$-dim, 25\,Hz} \\
Encoder strides    & \multicolumn{2}{c}{$[4,5,6,8]$} \\
Decoder strides    & $[5,4,4,3,2,2]$ & $[5,4,4,3,2,2,2]$ \\
Enc./dec. width    & \multicolumn{2}{c}{128\,/\,1536} \\
\midrule
Training data      & 600k tracks & 300k tracks \\
                   & 40k hours   & 20k hours \\
Updates            & \multicolumn{2}{c}{500k} \\
Segment / batch    & \multicolumn{2}{c}{4\,s / 32 (global)} \\
Hardware           & \multicolumn{2}{c}{8$\times$ A100 80GB, FP32} \\
Optimizer          & \multicolumn{2}{c}{AdamW, lr $10^{-4}$, 5k warmup} \\
Seed               & \multicolumn{2}{c}{42} \\
\bottomrule
\end{tabular}
\captionof{table}{MSVAE configuration.}
\label{tab:msvae-config}
\end{minipage}
\par\addvspace{7pt}

In Table~\ref{tab:loss-weights}, the two semantic streams use identical KL and alignment weights. Discriminator-specific weights are absorbed into the aggregate adversarial and feature-matching terms. The discriminator ensemble comprises MPD with periods $\{2,3,5,7,11\}$, MRD, and MSS-CQTD, and is active from the first update. In Stage~II, the KL and alignment terms are disabled because the encoder-side components are frozen, while the reconstruction and adversarial weights remain unchanged.

\par\addvspace{6pt}
\noindent\begin{minipage}{\columnwidth}
\setlength{\tabcolsep}{3pt}
\centering
\normalsize
\renewcommand{\arraystretch}{1.18}
\begin{tabular}{@{}llc@{}}
\toprule
Term & Symbol & Weight \\
\midrule
L/R multi-resolution STFT   & $\lambda_{\mathrm{stft}}$      & 10 \\
Stereo sum-difference STFT  & $\lambda_{\mathrm{stereo}}$    & 10 \\
Adversarial                 & $\lambda_{\mathrm{adv}}$       & 1.0 \\
Feature matching            & $\lambda_{\mathrm{fm}}$        & 1.5 \\
KL (both streams)           & $\lambda^{s}_{\mathrm{KL}}$    & 0.01 \\
Alignment (both streams)    & $\lambda^{s}_{\mathrm{align}}$ & 50 \\
\bottomrule
\end{tabular}
\captionof{table}{MSVAE training-loss weights.}
\label{tab:loss-weights}
\end{minipage}
\par\addvspace{7pt}

\subsection{Frozen Encoder Interfaces}
\label{app:frozen_interfaces}

Table~\ref{tab:refs} summarizes the frozen encoders and the feature layers used by MSVAE. SheetSage and FireRedASR features are extracted offline for semantic alignment, whereas the frozen MuQ encoder remains part of the MSVAE encoding path. MuQ processes the left and right channels independently, preserving channel-specific information in the semantic substrate. SheetSage melody and harmony states are indexed by beat-synchronous positions; we expand them to the 25-Hz latent grid by repeating each state in proportion to its timestamped duration and compensating accumulated rounding errors. FireRedASR operates at approximately 25~Hz, allowing the corresponding segment to be selected without temporal interpolation.

\par\addvspace{6pt}
\noindent\begin{minipage}{\columnwidth}
\setlength{\tabcolsep}{3pt}
\centering
\small
\renewcommand{\arraystretch}{1.08}
\begin{tabular}{@{}llcl@{}}
\toprule
Encoder & Layer used & Dim & Rate adaptation \\
\midrule
MuQ              & layer 6          & 1024 & native \\
SheetSage        & \shortstack[l]{melody +\\harmony} & 1024 & \shortstack[l]{beat-sync\\$\rightarrow$ 25\,Hz} \\
FireRedASR-AED-L & final encoder    & 1280 & native $\approx$25\,Hz \\
\bottomrule
\end{tabular}
\captionof{table}{Frozen encoder interfaces.}
\label{tab:refs}
\end{minipage}
\par\addvspace{7pt}

\subsection{Downstream Generator Configurations}
\label{app:downstream_config}

Table~\ref{tab:gen-config} summarizes the downstream generators. Within each task, all representations share the same backbone, training data, optimization schedule, number of updates, and inference configuration; only the input and output projections are adapted to the latent dimensionality, while sequence lengths follow each representation's native frame rate.

Both generators are non-autoregressive conditional flow-matching transformers with bidirectional SDPA attention, RoPE, and gated-SiLU feed-forward layers. They use AdamW with peak learning rate $10^{-4}$, $\beta=(0.9,0.999)$, weight decay $0.01$, gradient clipping $1.0$, FP16, and EMA decay $0.9999$ on 8$\times$ A100 80GB GPUs. Frame counts in Table~\ref{tab:gen-config} refer to MSVAE. For the duration analysis, separate 90- and 150-second generators (2250 and 3750 frames) are trained from scratch under the same recipe.

\par\addvspace{6pt}
\noindent\begin{minipage}{\columnwidth}
\setlength{\tabcolsep}{3pt}
\centering
\small
\renewcommand{\arraystretch}{1.08}
\begin{tabular}{@{}lcc@{}}
\toprule
 & Lyrics-to-song & Text-to-music \\
\midrule
Model scale      & 1B & 402M \\
Layers           & 16 & 16 \\
Hidden / FFN     & 2048\,/\,8192 & 1536\,/\,3072 \\
Heads $\times$ dim & 32$\times$64 & 32$\times$48 \\
Conditioning     & lyric + style (512) & MuLan (segment) \\
Segment (frames) & 45\,s (1125) & 30\,s (750) \\
\midrule
Updates          & 400k & 300k \\
Batch (global)   & 8$\times$8 (64) & 32$\times$8 (256) \\
LR schedule      & \multicolumn{2}{c}{warmup + linear decay} \\
\bottomrule
\end{tabular}
\captionof{table}{Downstream generator configurations.}
\label{tab:gen-config}
\end{minipage}
\par\addvspace{7pt}

\section{Additional Lyrics-to-Song Results}
\label{app:additional_l2s}

\subsection{Baselines with Untimed Lyric Conditioning}
\label{app:untimed_lyrics}

Table~\ref{tab:untimed_baselines} evaluates all representations under the same untimed lyric-conditioning protocol. Each generator receives the complete phoneme-token sequence as an untimed prefix followed by padding, without sentence-level timestamps or line-to-frame assignment; the architecture and training protocol are otherwise shared. The MSVAE row is repeated for direct comparison. PER can exceed one because insertion errors are counted. Bold and underlining mark the best and second-best results in the table.

\begin{table*}[!t]
\begingroup
\centering
\normalsize
\setlength{\tabcolsep}{4pt}
\renewcommand{\arraystretch}{1.0}
\begin{tabular}{@{}lccccccccc@{}}
\toprule
\multirow[c]{2}{*}{Representation} & \multicolumn{2}{c}{Audiobox Aesthetics} & \multicolumn{5}{c}{SongEval} & \multirow[c]{2}{*}{MuLan$\uparrow$} & \multirow[c]{2}{*}{PER$\downarrow$} \\
\cmidrule(lr){2-3}\cmidrule(lr){4-8}
& PQ$\uparrow$ & CE$\uparrow$ & Coh.$\uparrow$ & Mus.$\uparrow$ & Mem.$\uparrow$ & Cla.$\uparrow$ & Nat.$\uparrow$ & & \\
\midrule
Stable Audio Open & 7.86 & 7.38 & 3.10 & 2.87 & 2.92 & 2.84 & 2.79 & 0.593 & 0.906 \\
ACE-Step 1.5 & \underline{7.89} & \textbf{7.47} & \underline{3.35} & \underline{3.11} & \underline{3.15} & \underline{3.08} & \underline{3.01} & \underline{0.646} & \underline{0.851} \\
SAME-S & 7.29 & 7.10 & 2.82 & 2.52 & 2.54 & 2.49 & 2.51 & 0.557 & 1.579 \\
SAME-L & 7.57 & 7.28 & 3.16 & 2.88 & 2.92 & 2.82 & 2.84 & 0.585 & 0.877 \\
\midrule
MSVAE & \textbf{8.02} & \underline{7.44} & \textbf{3.60} & \textbf{3.39} & \textbf{3.46} & \textbf{3.38} & \textbf{3.26} & \textbf{0.678} & \textbf{0.136} \\
\bottomrule
\end{tabular}
\captionof{table}{Lyrics-to-song generation under untimed conditioning.}
\label{tab:untimed_baselines}
\endgroup
\par\vspace{5pt}
\begingroup
\centering
\normalsize
\setlength{\tabcolsep}{3.2pt}
\renewcommand{\arraystretch}{1.0}
\begin{tabular}{@{}lccccccccccc@{}}
\toprule
\multicolumn{3}{c}{Training configuration} & \multicolumn{2}{c}{Audiobox Aesthetics} & \multicolumn{5}{c}{SongEval} & \multirow[c]{2}{*}{MuLan$\uparrow$} & \multirow[c]{2}{*}{PER$\downarrow$} \\
\cmidrule(lr){1-3}\cmidrule(lr){4-5}\cmidrule(lr){6-10}
Model scale & Songs & Length & PQ$\uparrow$ & CE$\uparrow$ & Coh.$\uparrow$ & Mus.$\uparrow$ & Mem.$\uparrow$ & Cla.$\uparrow$ & Nat.$\uparrow$ & & \\
\midrule
300M & 200k & 45 s & 7.91 & 7.19 & 3.42 & 3.22 & 3.29 & 3.20 & 3.06 & 0.645 & 0.145 \\
1B & 200k & 45 s$^\dagger$ & 8.02 & 7.44 & 3.60 & 3.39 & 3.46 & 3.38 & 3.26 & 0.678 & 0.136 \\
\addlinespace
1B & 500k & 45 s & 8.01 & 7.48 & 3.79 & 3.57 & 3.63 & 3.55 & 3.39 & 0.693 & 0.174 \\
\addlinespace
1B & 200k & 90 s & 7.93 & 7.40 & 3.65 & 3.41 & 3.50 & 3.44 & 3.29 & 0.659 & 0.175 \\
1B & 200k & 150 s & 8.02 & 7.49 & 3.60 & 3.37 & 3.48 & 3.38 & 3.24 & 0.653 & 0.236 \\
\bottomrule
\end{tabular}
\captionof{table}{Robustness and scaling under untimed lyric conditioning.}
\label{tab:l2s_scaling}
\endgroup
\end{table*}

Although the baseline generators retain non-degenerate audio-quality and musicality scores, their PERs range from $0.851$ to $1.579$, showing that they do not reliably associate the supplied phoneme sequence with the generated vocals. In contrast, MSVAE attains a PER of $0.136$ together with the best PQ, all five SongEval scores, and MuLan. Because the generator architecture, training protocol, and conditioning format are shared across representations, the comparison isolates the effect of the latent representation.

\subsection{Robustness and Scaling Analysis}
\label{app:scaling}

Table~\ref{tab:l2s_scaling} tests whether timestamp-free lyric correspondence depends on the capacity of the 1B-scale generator, whether it extends beyond 45 seconds, and how it changes with additional training data. In the table, Model scale denotes the nominal generator-size category, Songs the number of training songs, and Length both the training and evaluation duration; the $\dagger$ symbol identifies the primary configuration. All models receive an untimed phoneme-token prefix and are trained for 400k updates without sentence-level timestamps or auxiliary generator-side alignment objectives. The duration-specific models are trained independently from scratch using duration-specific lyric segments drawn from the same pool of 200 evaluation songs.

The 300M-scale model retains a low PER of $0.145$, while increasing the model scale to 1B provides further consistent improvements. Expanding the training set from 200k to 500k songs improves all five SongEval dimensions and MuLan, although PER rises to $0.174$; the additional 300k songs come from a less-curated portion of the corpus. Separately trained 90- and 150-second models preserve musicality close to the 45-second model while obtaining PERs of $0.175$ and $0.236$, showing that the effect is not specific to the 45-second setting.

\section{Additional Ablation Results}
\label{app:additional_ablation}

\subsection{Acoustic-Only Posterior}
\label{app:acoustic_only}

Table~\ref{tab:acoustic_only_complete} reports the complete reconstruction and MARBLE results for the acoustic-only posterior. Both variants use the same convolutional acoustic VAE; the Lead+ASR variant applies partitioned SheetSage and FireRedASR alignment during VAE training. Bold marks the better result in each column. Partitioned Lead+ASR alignment improves all probing metrics while retaining comparable reconstruction performance.

\subsection{Representation-Side versus Generator-Side Guidance}
\label{app:repa}

Table~\ref{tab:guidance_complete} reports the complete results for the controlled comparison between representation-side and generator-side semantic guidance. The first three rows share the same unaligned acoustic-only VAE, with REPA applied only during generator training. The final row instead retrains the acoustic VAE with partitioned Lead+ASR alignment. All generators use the same untimed phoneme-token conditioning. Bold and underlining mark the best and second-best results in each column.

\begin{table*}[!t]
\begingroup
\centering
\normalsize
\setlength{\tabcolsep}{2pt}
\renewcommand{\arraystretch}{1.0}
\begin{tabular}{@{}lccccccccc@{}}
\toprule
\multirow[c]{2}{*}{Semantic alignment} & \multicolumn{4}{c}{Reconstruction} & \multicolumn{5}{c}{MARBLE probing} \\
\cmidrule(lr){2-5}\cmidrule(lr){6-10}
& PESQ$\uparrow$ & STOI$\uparrow$ & $\mathrm{STFT}_{\log}\downarrow$ & $\mathrm{Mel}_{\log}\downarrow$ & EMO $R^2_{\mathrm{V}}\uparrow$ & EMO $R^2_{\mathrm{A}}\uparrow$ & Chords Acc.$\uparrow$ & GS Acc.$\uparrow$ & Beat $F_1\uparrow$ \\
\midrule
None & \textbf{2.333} & \textbf{0.749} & 0.137 & \textbf{0.087} & 0.152 & 0.550 & 0.398 & 0.490 & 0.346 \\
Lead+ASR (part.) & 2.257 & 0.738 & \textbf{0.136} & 0.089 & \textbf{0.382} & \textbf{0.619} & \textbf{0.610} & \textbf{0.616} & \textbf{0.499} \\
\bottomrule
\end{tabular}
\captionof{table}{Acoustic-only posterior: reconstruction and MARBLE probing.}
\label{tab:acoustic_only_complete}
\endgroup
\par\vspace{4pt}
\begingroup
\centering
\normalsize
\setlength{\tabcolsep}{4pt}
\renewcommand{\arraystretch}{1.0}
\begin{tabular}{@{}lccccccccc@{}}
\toprule
\multirow[c]{2}{*}{Semantic guidance} & \multicolumn{2}{c}{Audiobox Aesthetics} & \multicolumn{5}{c}{SongEval} & \multirow[c]{2}{*}{MuLan$\uparrow$} & \multirow[c]{2}{*}{PER$\downarrow$} \\
\cmidrule(lr){2-3}\cmidrule(lr){4-8}
& PQ$\uparrow$ & CE$\uparrow$ & Coh.$\uparrow$ & Mus.$\uparrow$ & Mem.$\uparrow$ & Cla.$\uparrow$ & Nat.$\uparrow$ & & \\
\midrule
None & 7.50 & 7.14 & \underline{3.34} & \underline{3.09} & 3.11 & 3.00 & 2.97 & 0.636 & 1.394 \\
REPA (MERT+mHuBERT) & \underline{7.93} & \textbf{7.36} & 3.33 & 3.08 & \underline{3.14} & \underline{3.06} & \underline{2.99} & \underline{0.637} & 0.694 \\
REPA (Lead+ASR) & 7.63 & 7.18 & 3.02 & 2.76 & 2.83 & 2.76 & 2.72 & 0.599 & \underline{0.198} \\
Latent alignment (Lead+ASR) & \textbf{7.96} & \underline{7.35} & \textbf{3.45} & \textbf{3.28} & \textbf{3.32} & \textbf{3.26} & \textbf{3.12} & \textbf{0.662} & \textbf{0.182} \\
\bottomrule
\end{tabular}
\captionof{table}{Lyrics-to-song results under representation- and generator-side semantic guidance.}
\label{tab:guidance_complete}
\endgroup
\end{table*}

\section{Latent and Subspace Analysis}
\label{app:latent_analysis}

\subsection{Key Organization on GiantSteps}
\label{app:tsne}

In Table~\ref{tab:tsne-variants}, \emph{Gap} is the absolute difference between the two half-latent purities. For unaligned variants, columns 0--31 and 32--63 denote channel ranges only; under partitioned alignment, they correspond to $z^{\mathrm{comp}}$ and $z^{\mathrm{ling}}$. Both aligned rows use Lead+ASR: \emph{full-dim.} aligns both targets with all 64 channels, whereas \emph{partitioned} assigns them to disjoint 32-dimensional halves.

\begin{table}[!t]
\centering
\small
\setlength{\tabcolsep}{3pt}
\renewcommand{\arraystretch}{1.12}
\begin{tabular}{@{}llcccc@{}}
\toprule
Posterior & Alignment & Full & 0--31 & 32--63 & Gap \\
\midrule
Acoustic only  & None                 & 0.255 & 0.193 & 0.253 & 0.060 \\
Gated residual & None                 & 0.419 & 0.313 & 0.358 & 0.045 \\
Gated residual & Full-dim. & 0.486 & 0.436 & 0.379 & 0.057 \\
Gated residual & Partitioned & 0.498 & \textbf{0.555} & 0.076 & \textbf{0.479} \\
\bottomrule
\end{tabular}
\caption{Track-level GiantSteps key purity across controlled variants.}
\label{tab:tsne-variants}
\end{table}

This analysis uses one key annotation per track, allowing us to examine how the 24 musical keys are organized in the latent and which channels carry key-related compositional information. Each clip latent is averaged over time, and clips sharing the same \texttt{ori\_uid} are averaged again, so every track contributes one point (1{,}763 tracks). Table~\ref{tab:tsne-variants} reports leave-one-out 5-NN purity---the fraction of same-key neighbors---in the original latent space rather than on the two-dimensional embedding. Permuting labels 1{,}000 times with the neighbor graph fixed gives a chance level of $0.046$; the majority-class rate is $0.100$. Stars in the t-SNE panels denote $p\leq0.001$ under this permutation test. This neighborhood diagnostic is not directly comparable to supervised MARBLE probing; each t-SNE panel is fitted independently and used only for qualitative inspection.

\begin{figure*}[!p]
\centering
\textbf{(a) Acoustic-only posterior, no semantic alignment.}\par
\includegraphics[width=0.84\textwidth]{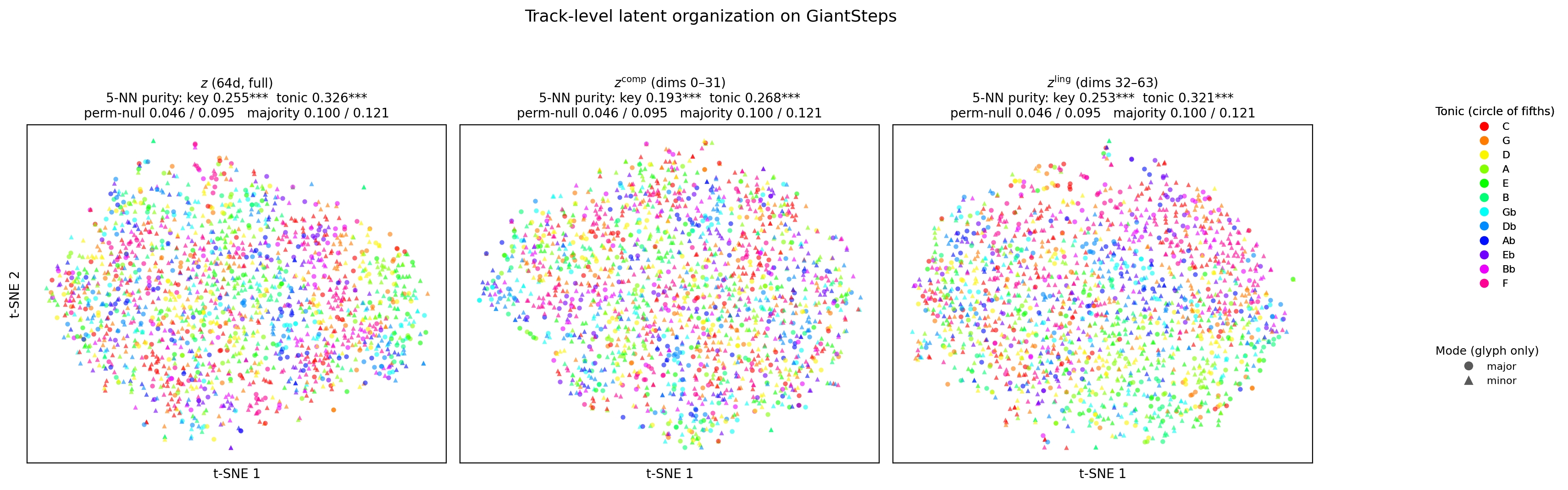}\par
\smallskip
\textbf{(b) Gated residual posterior, no semantic alignment.}\par
\includegraphics[width=0.84\textwidth]{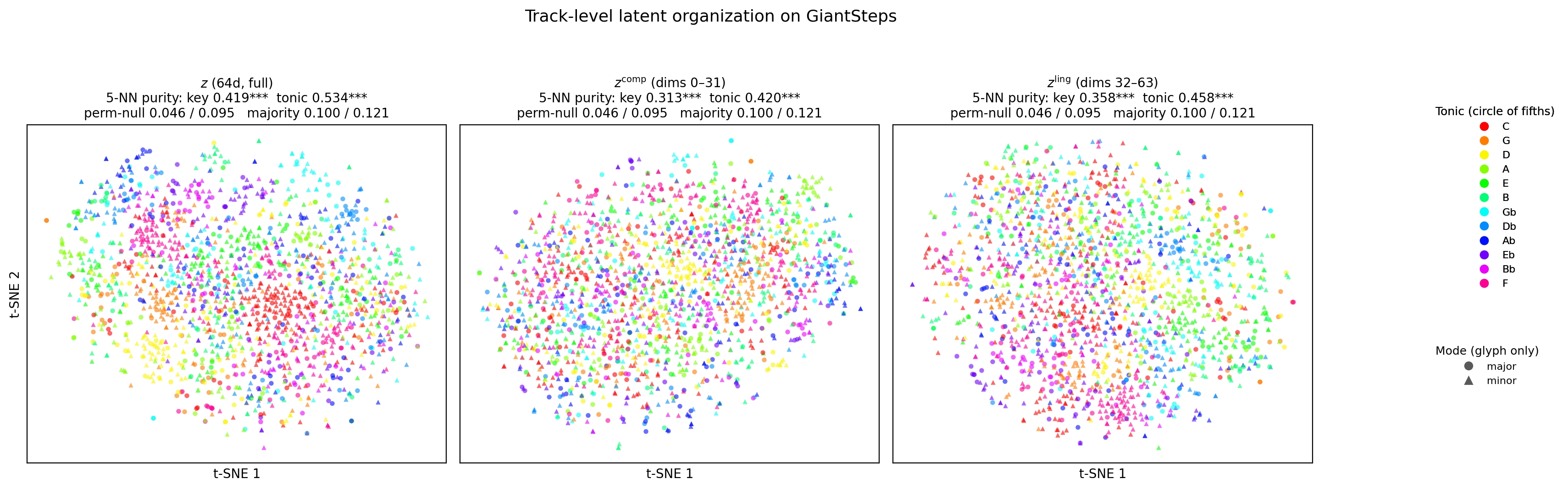}\par
\smallskip
\textbf{(c) Gated residual posterior with full-dimensional Lead+ASR alignment.}\par
\includegraphics[width=0.84\textwidth]{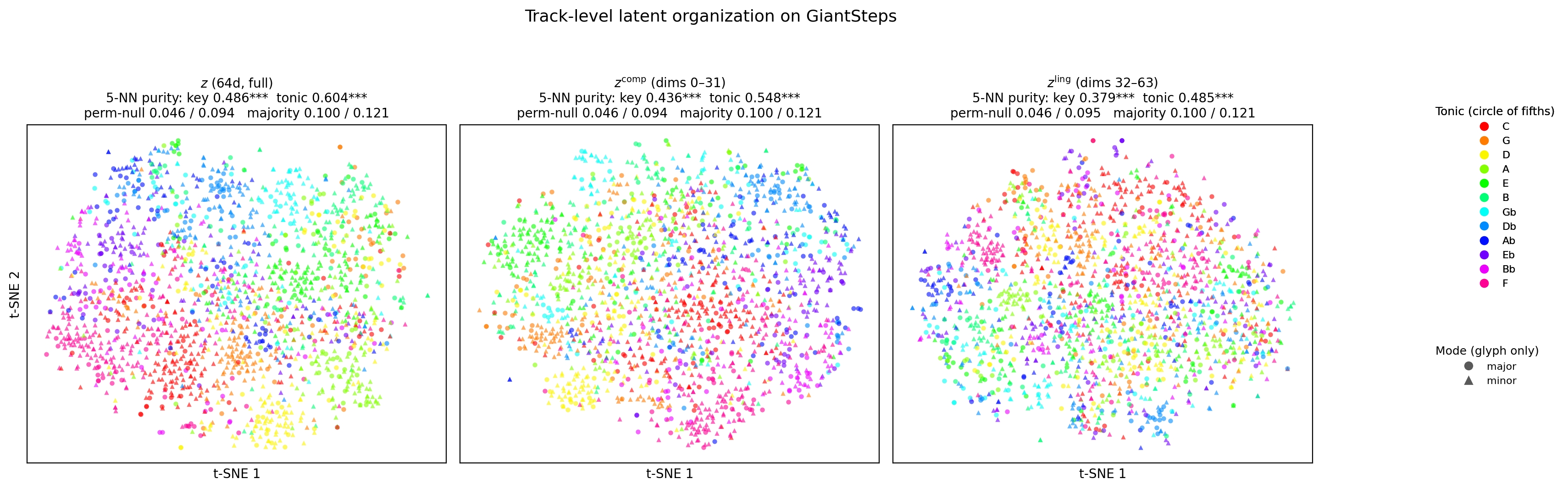}\par
\smallskip
\textbf{(d) MSVAE with partitioned Lead+ASR alignment.}\par
\includegraphics[width=0.84\textwidth]{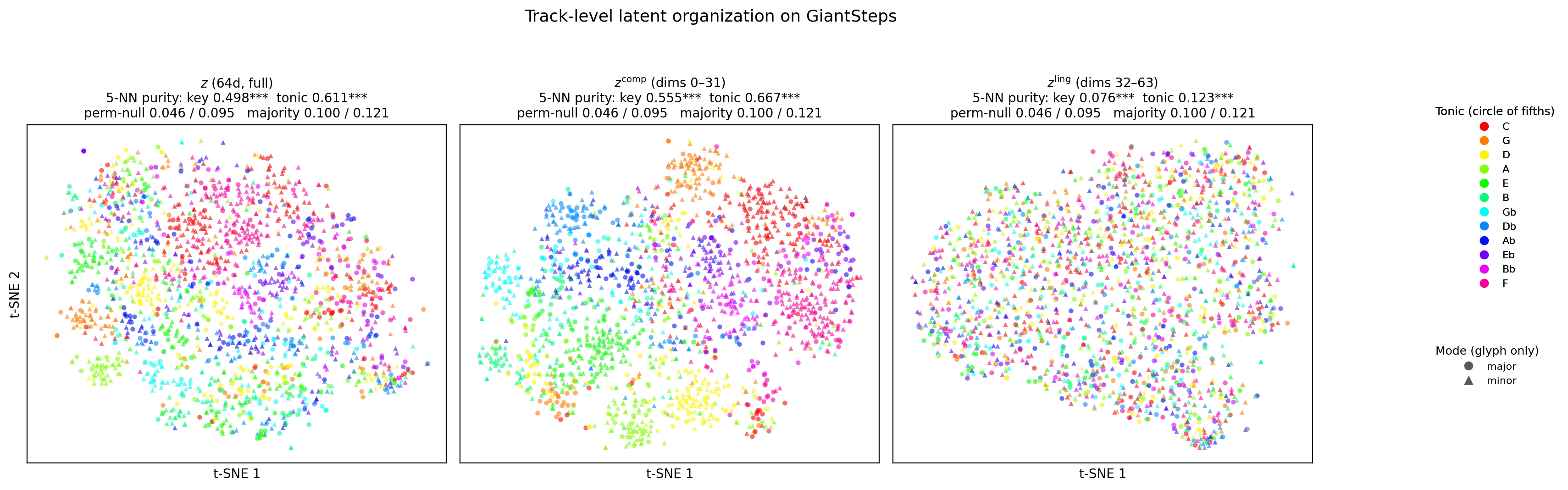}
\caption{Track-level GiantSteps t-SNE visualizations: (a) acoustic-only posterior, (b) gated residual without alignment, (c) gated residual with full-dimensional Lead+ASR alignment, and (d) MSVAE with partitioned Lead+ASR alignment. Each panel shows the full latent and its two halves.}
\label{fig:tsne-comparison}
\end{figure*}

Table~\ref{tab:tsne-variants} shows that key-related information concentrates in the designated compositional half only under partitioned alignment. The purity gap between the two halves reaches $0.479$, compared with $0.045$--$0.060$ for the other controlled variants. Equivalently, the ratio between the higher- and lower-purity halves is $7.30$, versus only $1.14$--$1.31$ for the other variants. Full-dimensional alignment raises full-latent purity but leaves the two halves similarly informative, indicating redundancy rather than specialization. This key-neighborhood analysis shows a strong concentration of key-related information in $z^{\mathrm{comp}}$ and a substantial reduction in $z^{\mathrm{ling}}$, although the latter remains above the permutation null. Because the analysis uses only key labels, it does not test whether $z^{\mathrm{ling}}$ carries linguistic-vocal information. Figure~\ref{fig:tsne-comparison} illustrates the same pattern: key-related structure remains distributed in the acoustic-only latent, while the two unaligned gated-residual halves differ only marginally and in the direction opposite to the partitioned model, consistent with their lack of assigned semantics.

Comparing the two Lead+ASR variants holds the posterior construction and semantic references fixed while changing how alignment is assigned to latent channels. Partitioning changes full-latent purity only from $0.486$ to $0.498$, yet raises the 0--31 score from $0.436$ to $0.555$ and lowers the 32--63 score from $0.379$ to $0.076$. The main difference is therefore where key-related neighborhood structure resides, rather than a large improvement in the full-code score. Under partitioned alignment, the compositional half alone even exceeds the full code on this key diagnostic; neither half does so under full-dimensional alignment. Reporting only whole-latent purity would miss this distinction.

\raggedbottom
\section{Limitations}
\label{app:limitations}

Our evaluation relies primarily on automated perceptual, semantic, transcription, and probing metrics, and we have not conducted a controlled human listening study. The downstream experiments use non-autoregressive conditional flow-matching transformers; whether the advantages of MSVAE transfer to autoregressive generators remains unverified. Stage~II improves the decoder to synthesize 48-kHz audio while retaining a frozen encoder that operates on 24-kHz input. The resulting 48-kHz output should therefore be interpreted as decoder-side bandwidth refinement rather than evidence that the latent representation captures information above the encoder's Nyquist frequency.

\end{document}